\documentclass{aa}
\usepackage[utf8]{inputenc}
\usepackage[T1]{fontenc}
\usepackage{graphicx}
\usepackage{color}
\usepackage[dvipsnames]{xcolor}
\usepackage{natbib}

\newcommand{\msun}{\ensuremath{\mathrel{\mathrm{M}_\odot}}}
\newcommand{\rsun}{\ensuremath{\mathrel{\mathrm{R}_\odot}}}
\newcommand{\myr}{\ensuremath{\mathrel{\mathrm{Myr}}}}
\newcommand{\gyr}{\ensuremath{\mathrel{\mathrm{Gyr}}}}
\newcommand{\pc}{\ensuremath{\mathrel{\mathrm{pc}}}}
\newcommand{\au}{\ensuremath{\mathrel{\mathrm{AU}}}}
\newcommand{\kpc}{\ensuremath{\mathrel{\mathrm{kpc}}}}
\newcommand{\ergs}{\ensuremath{\mathrel{\mathrm{erg}/\mathrm{s}}}}
\newcommand{\kms}{\ensuremath{\mathrel{\mathrm{km}/\mathrm{s}}}}
\newcommand{\msy}{\mathrel{\msun\;\mathrm{yr}^{-1}}}

\newcommand{\enulx}{\ensuremath{\mathbb{E}[\mathcal{N}_\mathrm{ULX}]}}
\newcommand{\npop}{\ensuremath{\mathrm{N}_\mathrm{pop}}}
\newcommand{\fracb}{\ensuremath{\mathrm{f}_\mathrm{bin}}}
\newcommand{\concpop}{\ensuremath{\mathrm{R}_\mathrm{h,2}/\mathrm{R}_\mathrm{h,1}}}
\newcommand{\rhmcl}{\ensuremath{\mathrm{Tidal}}}
\newcommand{\nf}{\texttt{NF}\xspace}
\newcommand{\rtid}{\ensuremath{\mathrm{R}_\mathrm{tid}}}  
\newcommand{\rgc}{\ensuremath{\mathrm{R}_\mathrm{gc}}}  
\newcommand{\rh}{\ensuremath{\mathrm{R}_\mathrm{h}}}  
\newcommand{\smt}{\ensuremath{\mathrm{M}_\mathrm{tot}}}  
\newcommand{\rc}{\ensuremath{\mathrm{R}_\mathrm{c}}}  
\newcommand{\xrc}{\ensuremath{\mathrm{M}_\mathrm{c}}}  
\newcommand{\roc}{\ensuremath{\rho_\mathrm{c}}}  
\newcommand{\threl}{\ensuremath{\mathrm{t}_\mathrm{h, rel}}}  

\newcommand{\lx}{\ensuremath{\mathrm{L}_\mathrm{X}}}
\newcommand{\ledd}{\ensuremath{\mathrm{L}_\mathrm{Edd}}}
\newcommand{\lxmax}{\ensuremath{\mathrm{L}_\mathrm{X, max}}}
\newcommand{\kacc}{\ensuremath{\mathrm{k}_\mathrm{acc}}}
\newcommand{\kdon}{\ensuremath{\mathrm{k}_\mathrm{don}}}
\newcommand{\macc}{\ensuremath{\mathrm{M}_\mathrm{acc}}}
\newcommand{\mdon}{\ensuremath{\mathrm{M}_\mathrm{don}}}
\newcommand{\dm}{\ensuremath{\dot{\mathrm{M}}}}
\newcommand{\dmedd}{\ensuremath{\dot{\mathrm{M}}_{\mathrm{Edd}}}}
\newcommand{\sep}{\ensuremath{a}}
\newcommand{\ecc}{\ensuremath{e}}
\newcommand{\tphys}{\ensuremath{\mathrm{t}_\mathrm{phys}}}
\newcommand{\tphysmin}{\ensuremath{\mathrm{t}_\mathrm{phys, min}}}
\newcommand{\tphysmax}{\ensuremath{\mathrm{t}_\mathrm{phys, max}}}
\newcommand{\dtulx}{\ensuremath{\Delta \mathrm{t}_\mathrm{ULX}}}

\newcommand{\nprog}{\ensuremath{\mathrm{N}_{\mathrm{prog}}}}
\newcommand{\med}[1]{\ensuremath{\mathrm{med}(#1)}}
\newcommand{\tulxstart}{\ensuremath{\med{\mathrm{t}_{\mathrm{ULX, start}}}}}
\newcommand{\enulxtotal}{\ensuremath{\enulx_{total}}}

\newcommand{\gbhms}{\ensuremath{G_{\textrm{BH,MS}}}}
\newcommand{\gbhhg}{\ensuremath{G_{\textrm{BH,HG}}}}
\newcommand{\gbhcheb}{\ensuremath{G_{\textrm{BH,CHeB}}}}
\newcommand{\gimbh}{\ensuremath{G_{\textrm{IMBH}}}}
\newcommand{\gns}{\ensuremath{G_{\textrm{NS}}}}

\newcommand{\mzams}{\ensuremath{\mathrm{M}_\mathrm{ZAMS}}}
\newcommand{\mcomp}{\ensuremath{\mathrm{M}_\mathrm{comp}}}
\newcommand{\fesc}{\ensuremath{\mathrm{f}_\mathrm{esc}}}
\newcommand{\fbin}{\ensuremath{\mathrm{f}_\mathrm{bin}}}
\newcommand{\fprim}{\ensuremath{\mathrm{f}_\mathrm{prim}}}
\newcommand{\fpris}{\ensuremath{\mathrm{f}_\mathrm{pris}}}

\titlerunning{Ultraluminous X-ray sources in Globular Clusters}
\authorrunning{Wiktorowicz et al.}

\begin{document}

\title{Ultraluminous X-ray sources in Globular Clusters}
\author{Grzegorz Wiktorowicz\inst{1}\thanks{E-mail: gwiktoro@camk.edu.pl},
        Mirek Giersz\inst{1},
        Abbas Askar\inst{1},
        Arkadiusz Hypki\inst{1,2},
        Lucas Helstrom\inst{1}}
\institute{  
     Nicolaus Copernicus Astronomical Center, Polish Academy of Sciences, Bartycka 18, 00-716 Warsaw, Poland\\
     \and
     Faculty of Mathematics and Computer Science, A. Mickiewicz University, Uniwersytetu Pozna\'nskiego 4, 61-614 Pozna\'n, Poland
}
\date{Accepted XXX. Received YYY; in original form ZZZ}

\abstract{

This paper investigates the formation, populations, and evolutionary paths of ultraluminous X-ray sources (ULXs) within globular clusters (GCs). ULXs, characterised by their extreme X-ray luminosities, present a challenge to our understanding of accretion physics and compact object formation. While previous studies have largely focused on field populations, this research examines the unique environment of GCs, where dynamical interactions play a significant role. Using the MOCCA Monte Carlo code, we explore how dynamics influences ULX populations within these dense stellar clusters.

Our findings reveal that dynamical processes, such as binary hardening and exchanges, can both facilitate and impede ULX formation in GCs. The study explores the impact of parameters including the initial binary fraction, tidal filling, and multiple stellar populations on the evolution of ULXs. We find that non-tidally filling clusters exhibit significantly larger ULX populations compared to tidally filling ones.

The results indicate that the apparent scarcity of ULXs in GCs may be related to the older stellar populations of GCs relative to the field. Furthermore, the study identifies a population of "escaper" ULXs, which originate in GCs but are ejected and emit X-rays outside the cluster. Our simulations reveal that these escapers constitute about one-seventh of the total ULX population. However, for neutron star accretors specifically, escapers are twice as common as their in-cluster counterparts. Notably, only $4\%$ of in-cluster ULXs contain neutron star accretors. These escapers may significantly contribute to the observed field ULX population.}


\maketitle

\section{Introduction}

Ultraluminous X-ray sources (ULXs) have been a subject of intense study in the field of high-energy astrophysics for several decades. These enigmatic objects, characterized by X-ray luminosities exceeding $10^{39}\ergs$, challenge our understanding of accretion physics and compact object formation \citep[see e.g.,][for a recent review]{Kaaret1708,King2306}. With over $1800$ known ULXs identified to date \citep{Walton2201}, they represent a significant population of extreme X-ray emitters in the Universe, yet their fundamental nature remains a topic of ongoing debate.

The diversity of ULXs has been highlighted in recent classification schemes \citep[e.g.,][]{Wiktorowicz1709}, which attempt to categorize these objects based on their observational properties and potential underlying physical mechanisms. While the majority of ULXs were initially thought to harbor intermediate-mass black holes (IMBHs), our understanding has evolved considerably with the detection of pulsar accretors \citep{Bachetti1410}. 

The importance of studying ULXs, particularly in the context of globular clusters (GC), cannot be overstated. ULXs serve as laboratories for extreme accretion physics, potentially shedding light on the formation and evolution of double compact objects - the progenitors of gravitational wave sources. Moreover, ULXs continue to be considered as candidates for elusive IMBHs with masses in the range of $10^{2}$ to up to $10^{5}\msun$ \citep[and references therein]{Kaaret1708}. The existence of IMBHs may provide crucial insights into the formation pathways of supermassive BHs observed at the centers of most galaxies. These intermediate-mass objects could also explain the presence of supermassive BHs observed at high redshift \citep{Greene2008,Askar23}.

In this paper, we present a comprehensive study of ULXs in GCs, focusing on their stellar-mass manifestations. The dense stellar environment introduces additional complexities and possibilities for exotic binary formations. By examining their properties, distribution, and potential formation mechanisms, we aim to contribute to the broader understanding of ULX physics and their role in compact object evolution.

Dense, quasi-spherical collections of stars known as GCs orbit the centers of galaxies. These typically old stellar systems, often containing millions of stars within a relatively small volume, are hubs of dynamical interactions. The masses of Milky Way GCs range from a few $10^{3}$ to a few $10^{6} \ \msun$, with half-mass radii typically between 1.5 pc and a few tens of pc, and ages spanning from 8 to 13.5 Gyr \citep{Harris9610,Salaris0206,VandenBerg1310,Baumgardt1808}. Detailed properties of specific Galactic GCs can be found in the catalogs of \citet[][updated 2010]{Harris9610}\footnote{\url{https://physics.mcmaster.ca/~harris/mwgc.dat}} and \citet{Baumgardt1808}\footnote{\url{https://people.smp.uq.edu.au/HolgerBaumgardt/globular/parameter.html}}.

The evolution of GCs is governed by the relaxation process connected with continuous distant gravitational interactions between all objects in the cluster. This leads to global phenomena like core collapse, that is contraction of the cluster core and gradual increase in central density, and mass segregation, that is the tendency of heavier stars to occupy the central regions while pushing lighter stars outwards. Close dynamical interactions also influence stars and binaries individually, affecting binary parameters (e.g., separation and eccentricity), or through exchanges (e.g., binary changes one of it's component into another star), ejections (star or binary becomes unbound with the cluster), and binary formation (single stars became bound and form a binary).

Such dynamical encounters in dense star clusters can lead to the formation of close binary systems, including X-ray binaries \citep{Pooley0307,Ivanova0503,Ivanova0805,Kremer1801,Kremer1901}, cataclysmic variables \citep{Oh2407,Belloni1902}, and gravitational wave progenitors \citep[e.g.,][and references therein]{Benacquista1312,Mandel2212}. Recent studies have emphasized the importance of GC dynamics in shaping exotic stellar populations \citep{DiCarlo2111,Oh2407}, highlighting the role of these environments in producing and evolving compact object binaries.

While the majority of ULXs are detected in star-forming regions of galaxies, there is increasing evidence for their presence in GCs as well. These GCULXs are found exclusively in extragalactic GCs, all of which are unresolved at distances of >16 Mpc, with no ULXs detected in Milky Way GCs to date. Pioneering observations of GCULXs have been reported in several galaxies, with varied GC environments, spectral behavior, and variability patterns.

\citet{Maccarone07} discovered the first ULX in a GC (NGC 4472). It exhibits an X-ray luminosity of $4 \times 10^{39}\ergs$ and rapid variability. Similar detections were reported in NGC 1399 \citep{Shih10,Irwin10}, with more GCULXs in NGC 4472 \citep{Maccarone11} and NGC 4649 \citep{Roberts12}. Some sources identified in GCs exhibited flares just above the ULX-defining limit of $10^{39}\ergs$ \citep{Irwin16,Sivakoff05}.

Recent observations have extended the known GCULX population, including seven additional GCULXs associated with M87 GCs \citep{Dage20}, three in NGC 1316 \citep{Dage21}, $10$ new candidates in GCs of massive ($>10^{11.5} M_{\odot}$) early-type galaxies \citep{Thygesen2301}, and two more in NGC 4261 \citep{Nair23}. The current population of approximately 30 known GCULXs, identified through systematic searches of thousands of GC systems in elliptical galaxies, represents a remarkably rare phenomenon compared to the $\sim1800$ ULXs identified in other galactic environments \citep{Walton2201}.

In dense stellar systems, compact objects can be readily ejected, which can considerably impact the formation channels of such objects \citep[e.g.,][]{Kulkarni9307}. Furthermore, GCs typically exhibit high ages and, consequently, fewer higher-mass stars. Given their presence in older stellar populations, GCULXs are likely powered by low-mass X-ray binaries, in contrast to field ULXs which are typically associated with massive donor stars but often lack optical counterparts. Given their presence in older stellar populations, GCULXs are likely powered by low-mass X-ray binaries, in contrast to field ULXs which are expected to be also associated with low-mass as well as massive donor stars, but often lack optical counterparts. These characteristics contribute to the intriguing nature of GCULXs.

\section{Methodology}

\subsection{MOCCA Monte Carlo code}

The MOCCA \citep[MOnte Carlo Cluster evolution Code;][]{Hypki1302} is a state-of-the-art numerical tool that combines the Monte Carlo method for stellar dynamics \citep{Henon7110,Stodolkiewicz8601} with detailed stellar and binary evolution algorithms from the Binary Star Evolution code \citep[BSE;][]{Hurley0007,Hurley0202} with recent updates \citep[level-C from][and references therein]{Kamlah2204} and low-N scattering code FEWBODY \citep{Fregeau0407} to follow close dynamical interactions. This hybrid approach makes MOCCA particularly well-suited for studying the formation and evolution of exotic objects like ULXs in dense stellar environments, offering an optimal balance between computational efficiency and physical accuracy.

Recent improvements to the MOCCA code \citep{Hypki2212,Hypki2501,Giersz2411} have significantly enhanced its capabilities for ULX studies. First of all, the mass transfer calculations now incorporate a more sophisticated treatment of super-Eddington accretion, crucial for modeling ULX systems \citep{Shakura7301,Wiktorowicz1509}. Additionally, improvements in core radii calculations improved the Roche-lobe overflow calculations. The introduction of detailed evolutionary tracking provide unprecedented insight into the spatial distribution and evolutionary pathways of ULX systems within clusters equivalent to detailed outputs from codes like \texttt{startrack} \citep{Belczynski0801}.

A major advancement relevant to this study is MOCCA's capability to handle multiple stellar populations, a phenomenon now recognized as ubiquitous in GCs \citep[e.g.,][]{Piotto1503,Renzini1512,Gratton1911,Bastian1809,Milone2206}. While a comprehensive analysis of multiple stellar populations in our models is presented in \citet{Hypki2212,Hypki2501,Giersz2411}, here we specifically investigate their impact on ULX formation and evolution. The code now tracks different stellar populations with distinct chemical compositions, ages, and spatial distributions, allowing us to examine how properties specific to multiple stellar populations affect compact object formation and ULX characteristics. For detailed analysis of simulational results on multiple stellar populations and all recent changes to the MOCCA code see \citet[and references therein]{Giersz2411}.

The X-ray luminosity (\lx) of accretion disks is calculated using the Shakura-Sunyaev model \citep{Shakura7301}, accounting for both sub-Eddington and super-Eddington accretion regimes. The Eddington mass accretion rate is defined as $\dmedd = 2.2 \times 10^{-8} \macc \eta_{0.1}^{-1}\;[\msy]$, where \macc\ is the accretor mass and $\eta\;(=10\eta_{0.1})$ is the accretion efficiency. We assumed $\eta_{0.1}=1$. The Eddington luminosity is then $\ledd \approx 5.66 \times 10^{45}\;\dmedd\; [\text{erg s}^{-1}]$. The X-ray luminosity is calculated as:
\begin{equation}
\lx = 
\begin{cases}
\ledd (\dm / \dmedd), & \text{for } \dm \leq \dmedd \\
\ledd [1 + \ln(\dm / \dmedd)], & \text{for } \dm > \dmedd
\end{cases}
\end{equation}
where $\dm$ is the mass accretion rate. See \citet{Lasota2312} for recent discussion of this formulation.


\subsection{Simulations}

\begin{figure}
    \centering
    \includegraphics[width=\linewidth]{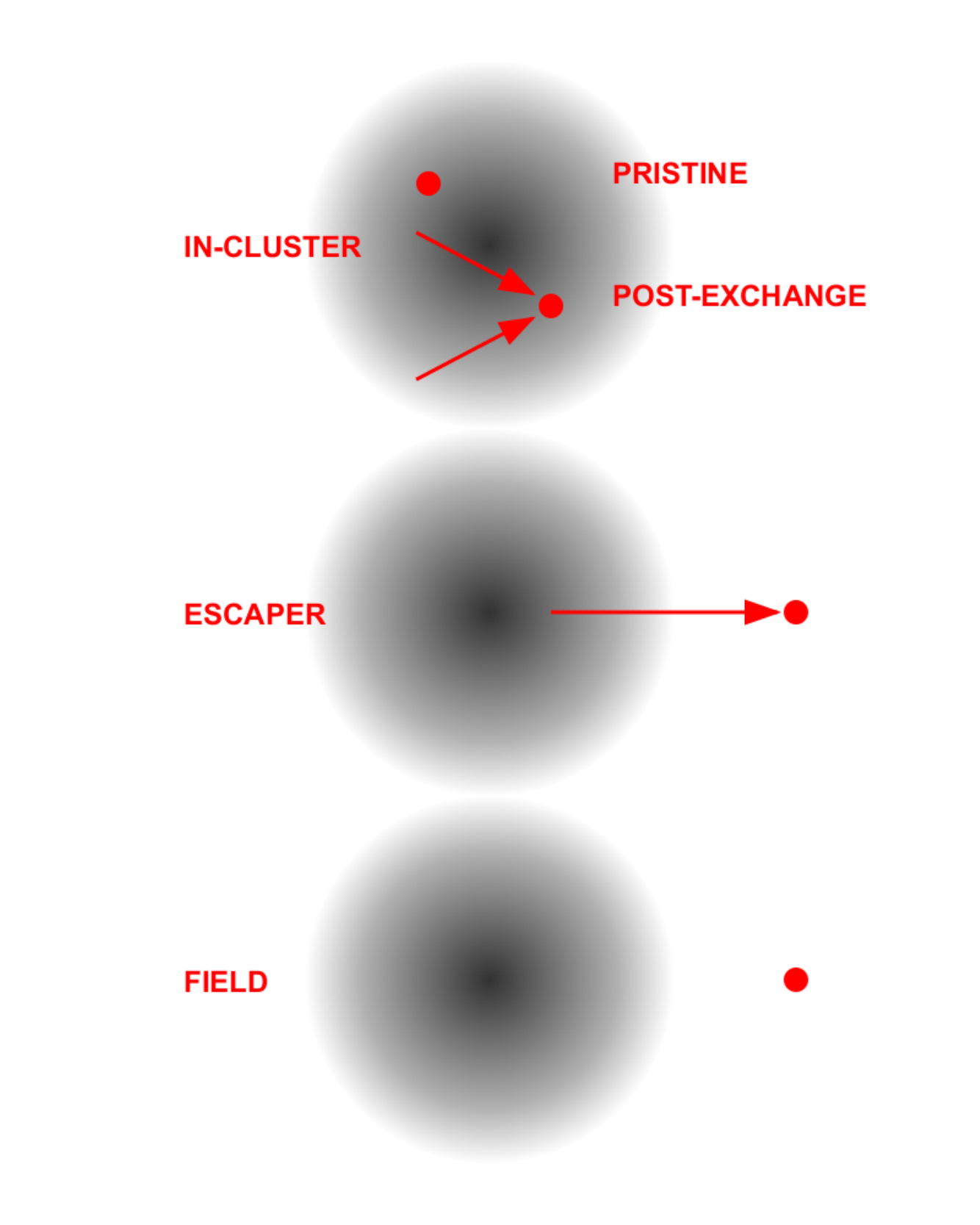}
    \caption{Schematic representation of the origins and eventual locations of ULX progenitors and ULXs in GCs. Progenitors can either form and remain bound to the cluster, leading to in-cluster ULX emission ("IN-CLUSTER"), or be dynamically ejected and emit as ULXs outside the cluster ("ESCAPER"). Additionally, ULXs originating and emitting in the field ("FIELD") represent progenitors formed without direct interaction with cluster dynamics.}
    \label{fig:gculx_formation}
\end{figure}

Figure \ref{fig:gculx_formation} illustrates three pathways for ULX formation in relation to GCs. In the "IN-CLUSTER" scenario, ULX progenitors form and remain bound to the cluster, often through multiple dynamical interactions, including exchanges. "ESCAPER" ULXs originate in the GC but are ejected through interactions, emitting outside the cluster while retaining their GC-imprinted dynamical history. "FIELD" ULXs, by contrast, form and evolve in isolation. The key distinction lies in the dynamical imprint: "ESCAPER" ULXs bear the hallmarks of GC interactions, unlike "FIELD" systems, which provide a baseline for understanding the unique contribution of clusters to ULX formation.

The MOCCA code employs hundreds of parameters to govern the dynamical and stellar evolution of GCs. Based on previous studies of compact object populations - including cataclysmic variables \citep{Belloni1706}, ULXs in the field \citep{Wiktorowicz1904}, and double white dwarf systems \citep{Hellstrom2410} - we identified key parameters that most significantly influence ULX formation and evolution. Our parameter selection strategy focused on those that control dynamics and initial stellar population properties. This systematic approach allows us to explore the most relevant parameter space while maintaining computational feasibility. In this section we present discussion of the utilized parameters.

In all simulations we adopted the \citet{Kroupa0104} initial mass function (IMF) for both stellar populations, with mass ranges of $0.08$--$150\msun$ for the first population and $0.08$--$20\msun$ for the second population. For binary systems, we implemented a pairing mechanism, that combines a uniform mass ratio distribution ($0.1 < q < 1.0$) for massive stars ($M>5\msun$) following \citet{Kiminki1205,Sana1207,Kobulnicky1408}, with random pairing for lower-mass stars. All clusters were initialized in virial equilibrium ($Q_{\rm vir}=0.5$). For the underlying density distribution, we used King models with concentration parameters of $W_0=3.0$ and $W_0=7.0$ for the first and second populations, respectively, representing moderately concentrated initial configurations. More details on the MOCCA parameters can be found in \citet{Kamlah2204,Hypki2212,Hypki2501,Giersz2411}.

For all simulations with dynamics enabled, we tracked the evolution of escaped systems to analyze the properties of their populations and potential ULXs that can form after their progenitors are ejected from the cluster. These escapers evolve as isolated binaries, similar to field binary evolution \citep[e.g.,][]{Fragos1503,Wiktorowicz1904,Zuo2105}. Systems can escape the cluster through multiple mechanisms: (1) dynamical interactions, particularly strong few-body encounters, (2) natal kicks from supernovae explosions, including both direct and Blaauw kicks, (3) gradual two-body relaxation \citep{Spitzer7103}, and (4) tidal stripping of outer cluster regions by the host galaxy's gravitational field \citep{Gnedin9701}.

The galactocentric distance ($\rgc$) for each simulation was scaled to maintain an initial tidal radius of $\rtid\approx43\pc$ across all models. This normalization of the tidal radius facilitates meaningful comparisons between simulations, particularly when analyzing the degree of tidal filling and cluster structural parameters. The tidal radius follows the relation $\rtid \propto \rgc^{2/3}\smt^{1/3}$ \citep[e.g.,][]{Webb1302}, where \smt\ is the cluster mass. This scaling ensures that clusters experience comparable relative tidal forces despite different masses and orbital parameters. We note that while the initial tidal radii are identical, the clusters' subsequent evolution may lead to different filling factors\footnote{filling factor is \rtid/\rh, which is initially $\sim43$ for non-tidally filling simulations and $\sim3.6$ for tidally filling ones.} depending on their internal dynamics and mass-loss history.

We performed a comprehensive set of simulations exploring different initial conditions and physical parameters. Table \ref{tab:simulations} summarizes our simulation grid, where each model is identified by a unique label used throughout this paper. The simulations vary in the number of stellar populations ($\npop$), with varied spatial distribution of different generations through the concentration parameter ($\concpop$), defined as the ratio between the half-mass radii of subsequent generation relative to the first generation. The initial binary fraction ($\fracb$) was either $10\%$, or $95\%$, which affected also other parameters (see below). We explored both tidally filling (TF) and non-tidally filling (nTF) configurations ($\rhmcl$). Some simulations incorporate the new features (\nf) described below, particularly relevant for multiple population scenarios. For each simulation, we provide key structural parameters at $t=0\myr$, including the galactocentric radius ($\rgc$), half-mass radius ($\rh$), total stellar mass ($\smt$), core radius \citep[$\rc$; according to ][]{Casertano8511}, core mass ($\xrc$, i.e., mass inside $\rc$), and central density \citep[$\roc$; according to][]{Casertano8511}. Additionally, we computed corresponding field populations with dynamics turned off (denoted by \texttt{nodyn} in the model labels) to serve as control cases, though these are not shown separately in the table.

To better compare simulations with different binary fractions, we made several adjustments to the parameter space. Since a lower binary fraction results in lower total stellar mass (as binary systems typically have higher masses than single stars), we adjusted the initial number of objects to have a similar half-mass relaxation time for both models: simulations with $\fracb=95\%$ used $n=600,000$ objects, while those with $\fracb=10\%$ used $n=1,063,635$ objects. We also modified the initial distribution of semi-major axes. 
For simulations with $\fracb=95\%$, we used the modified version of the \citet{Kroupa9512} period distribution \citep[see][]{Belloni1706} for stars with $M<5\msun$, and the distributions from \citet{Sana1207,Oh1506} for stars with $M>5\msun$. For simulations with $\fracb=10\%$, we employed a log-uniform distribution for $M<5\msun$ and the \citet{Sana1207} period distribution for $M>5\msun$. Additionally, for the low binary fraction simulations, we disabled eigenevolution \citep{Belloni1706}. The separation was limited to the maximal value of $100\au$

In a part of our simulations (marked with NF in label), we introduced \nf to investigate the influence of multiple stellar populations \citep[see][for details]{Giersz2411}. Specifically, we implemented a delay time ($t_{\rm delay}$) of $100\myr$ for the second population, during which these stars act as gas particles without undergoing stellar evolution, relaxation, or dynamical interactions. The gas accumulation for the second population begins at $t_{\rm delay\_fraction}=0.5$, meaning halfway through the delay period, coinciding with the asymptotic giant branch (AGB) wind contribution phase. To accommodate this, we activated AGB wind accumulation for the second population formation. 

The cluster's orbital evolution was modified by implementing an artificial orbit change at $t=1000\myr$, where the galactocentric distance increases by a factor of 2 \citep[see][for details]{Giersz2411}. Finally, we employed single-population scaling, where the scaling parameters are applied exclusively to the first population, while the second population's velocities are independently normalized. First and second populations are separately in the virial equilibrium.

Except dynamical evolution parameters, our simulations incorporate several key prescriptions for stars and binaries, some of them particularly relevant to ULX formation, specifically: We employed the rapid supernova mechanism as described by \citet{Fryer1204} for both neutron star (NS) and black hole (BH) formation; Neutron star and BH natal kicks were drawn from a Maxwellian distribution with $\sigma=265\kms$ \citep{Hobbs0507}; For BHs, these kicks were scaled by the fallback factor, which depends on the supernova prescription; Neutron stars formed through electron-capture supernovae were assigned substantially lower kicks with $\sigma=3\kms$; Common envelope evolution was modeled using the parameters from the BSE code: $\alpha=0.5$ and $\lambda=0.0$. All simulated models in this study were set to a fixed metallicity of $\mathrm{Z}=0.001$, which represents the typical metallicity of Galactic globular clusters\footnote{see the Harris catalog of Milky Way globular clusters \url{https://physics.mcmaster.ca/~harris/mwgc.dat} \citep[2010 update]{Harris9610}.}.

\section{Results}

ULXs in this study are defined as binary systems with neutron star or black hole accretors exhibiting X-ray luminosities exceeding $10^{39}\text{ erg s}^{-1}$. White dwarf-powered ULXs are excluded from this analysis.

\subsection{Number evolution of ULXs}\label{sec:n_ulx_sims} 

\begin{figure}
    \centering
    \includegraphics[width=1\linewidth]{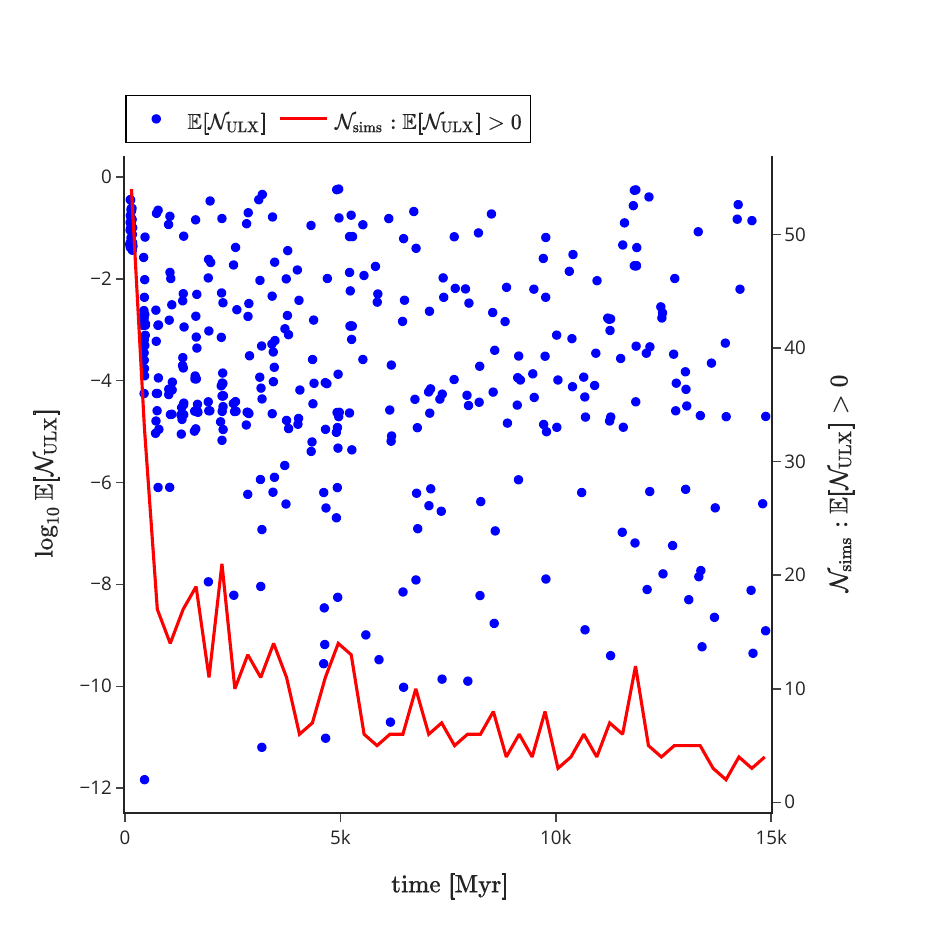}
    \caption{Evolution of ULX numbers in GCs. Expected number of ULXs (\enulx) as a function of time since GC formation. All simulations with non-zero expected rates are presented. The line indicates the number of simulations with non-zero predictions (i.e., number of dots in this time bin). Each time bin spans $300\;\myr$ and the points are centered on these bins.
    }
    \label{fig:n_ulx_sims}
\end{figure}

Figure \ref{fig:n_ulx_sims} illustrates the temporal evolution of the expected number of ULXs\footnote{Defined as the number of ULXs weighted by their observational probability in specific evolutionary time ranges. For example, an expected value of $0.001$ implies that, on average, one active ULX is observed across $1000$ simulations. Conversely, an expected value of $10$ indicates that, on average, $10$ ULXs are observed in a single simulation. This metric provides a probabilistic interpretation of ULX occurrence across simulations, accounting for their likelihood within evolutionary contexts.} (\enulx) across all simulations, accompanied by the count of simulations yielding non-zero \enulx\ for each time bin. The highest \enulx\ is observed in the initial evolutionary stages, typically within the first $300\myr$ (first bin in Figure \ref{fig:n_ulx_sims}). This trend persists across all simulations, irrespective of parameter variations. These findings align with previous studies on field populations, where dynamical interactions are negligible, which reported that the ULX phase predominantly occurs in the early evolutionary stages of binaries \citep[e.g.,][]{Wiktorowicz1709}. Observational evidence also supports this conclusion \citep[e.g.,][]{Wolter1808}.

This result suggests either:
\begin{itemize}
    \item Early ULX evolution is largely independent of cluster dynamics parameters, or
    \item Insufficient time has elapsed for dynamics to significantly influence ULX formation.
\end{itemize}

The number of simulations with $\enulx \neq 0$ decreases over time, resembling predictions for populations formed in burst-like star formation episodes \citep[e.g.,][]{Wiktorowicz1709}. In GCs, this trend may result from a combination of system age effects on ULX formation (as observed in field populations) and the interplay of positive and negative effects related to dynamics. Generally, our results indicate a higher likelihood of observing ULXs in younger stellar clusters ($\lesssim300\myr$) compared to older ones, which is consistent with observational results \citep[e.g.,][]{Dage2501}.

Figure \ref{fig:n_ulx_sims} reveals substantial variation in \enulx\ between simulations, particularly in later evolutionary phases. This variability likely stems from stochastic processes (simulation independent, that is resulting from random number generator seed) and varying dynamical effects (simulation dependent). The high variability in later evolutionary phases in comparison to the first bin ($0$ -- $300\myr$) where low variation is observed suggest that, \enulx\ is affected by dynamics in at least a significant way.

The variations in the otherwise monotonically decreasing number of non-zero predictions can be attributed to randomness. The peak at $5\gyr$ and even more pronounced peak at $12\gyr$ are not correlated to any significant changes in the GC structure and are not statistically significant, therefore can be treated as anomalies.

Please note that the provided \enulx\ values represent predictions per cluster. To derive observational expectations, one must sum these values across all observed clusters and account for observational limitations, such as magnitude limits and resolution constraints. A detailed discussion of observational predictions will be addressed in a separate study.

\subsection{Temporal evolution and parameter dependence of \enulx}

\begin{figure*}
    \centering
    \includegraphics[width=\textwidth]{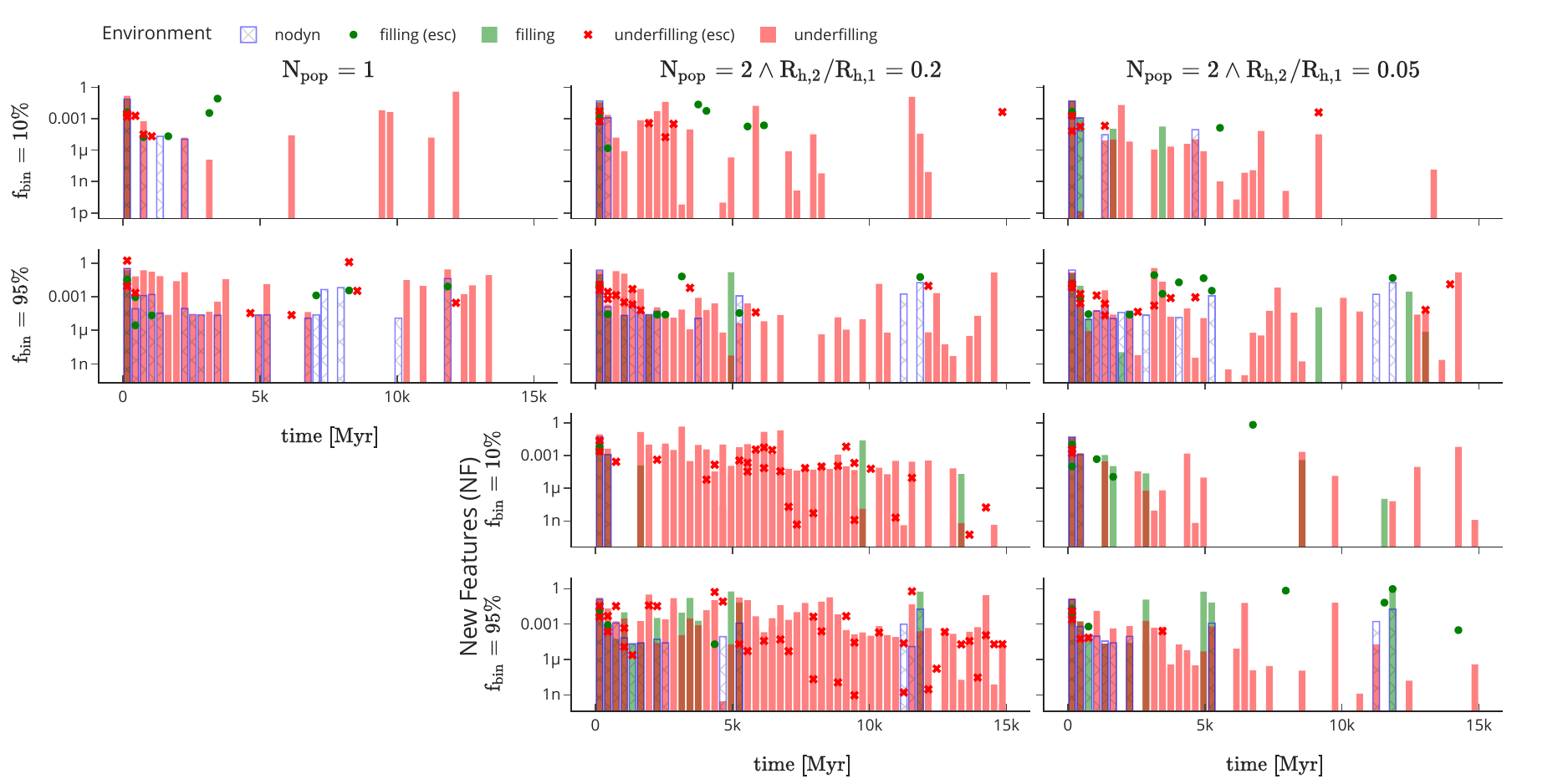}
    \caption{Comprehensive overview of the \enulx\ evolution since the formation of a GC. Each panel represents a set of simulations with identical parameters: number of populations (\npop), concentration (\concpop), binary fraction (\fracb), and the presence of new features (\nf). Within each panel, ULX populations from various environments are presented: tidally filling (TF) GC, non-tidally filling (nTF) GC, nodyn (binaries evolving without dynamical interactions), and escapers (progenitors formed in a GC, but experiencing the ULX phase after ejection from the cluster). See Table \ref{tab:simulations} for simulations parameters overview. Bars and points represent the \enulx\ in $300\myr$ bins}
    \label{fig:n_ulx_grid}
\end{figure*}

Figure \ref{fig:n_ulx_grid} presents a comprehensive analysis of \enulx\ dependence on model parameters and its temporal evolution. The most significant effects are observed in parameters related to the environment, specifically whether the GC is tidally filling (TF) or non-tidally filling (nTF), and the general impact of dynamics (in-cluster vs. field formation). These results support the hypothesis that dynamical interactions strongly influence the formation of ULXs in dense stellar environments. In particular, the exchange process may lead to the formation of binaries that are not possible through regular formation channels. The specific trends and patterns are described below.

Non-tidally filling clusters exhibit significantly larger ULX populations compared to tidally filling clusters. This discrepancy can be attributed to the extended time frame density available for dynamical interactions to enhance progenitor formation in non-tidally filling clusters, whereas tidally filling clusters experience more early ejections. More generally, the expected number of ULXs formed in GCs is substantially higher than in corresponding field populations, particularly for non-tidally filling clusters. Typically, ULXs formed in GCs demonstrate higher \enulx\ rates in specific time bins, and these rates are more continuous, whereas field populations mostly exhibit isolated occurrences throughout the timeline. These relations do not apply to early evolutionary stages where all simulations show high and similar values of \enulx\ (see Section \ref{sec:n_ulx_sims}).

Simulations with two populations (middle column) generally produce more ULXs than those with a single population (left column). This can be attributed to the higher concentration of the second population, which enhances ULX progenitor formation. However, when the concentration is excessively high (right column), this effect is less pronounced, which supports the claim that too frequent and strong dynamical interactions can substantially destroy ULX progenitors.

The initial binary fraction appears to have a limited influence on ULX formation in GCs. For $\fracb=95\%$ wide binaries are quickly destroyed in dynamical interactions and mostly remain binaries similar to ones for $\fracb=10\%$. The population of ULX progenitors remains sparse and emerges regardless of the initial relative number of binaries. Dynamical effects effectively destroy existing binaries and create new ones, a process primarily responsible for the formation of ULX progenitors at later evolutionary times. Conversely, the binary fraction influences field populations, with a positive correlation between initial binaries and ULX progenitors, as there are no external processes to destroy ULX progenitors or create new ones in isolation.

The incorporation of \nf has a noticeable effect on two-population simulations with moderate concentration ($\concpop=0.2$; Figure \ref{fig:n_ulx_grid}, middle column). Simulations with \nf exhibit higher \enulx\ values compared to those without \nf. The \nf do not affect field populations as they primarily relate to the time delay for the second population.

The escaper populations generally align with the ULX population formed in situ, suggesting ongoing ejection (relaxation and interactions), and thus a continuous supply of ULX progenitors to the escaper population, which later become ULXs while unbound to the cluster. There are generally fewer ULXs among escapers than in in-cluster populations. However, ULXs from escapers can dominate field populations. In general, there are no significant differences in the formation time of the ULXs from escapers to those in the field. An exception occurs in simulations with relatively strong ULX formation ($\npop=2$, $\concpop=0.2$), where ULXs from escapers form at evolutionary ages not covered by field populations. For tidally filling clusters, the rate of ULX formation from escapers can exceed that of in-cluster formation, especially in simulations without \nf.

\subsection{Formation of ULXs}

This section focuses on ULX progenitors - systems that will evolve into ULXs - and their evolution up to the initial ULX phase. It is important to note that some systems may experience multiple ULX phases interspersed with periods of quiescence or X-ray binary phases ($\lx<10^{39}\ergs$). These phase transitions typically occur on timescales of at least hundreds of years, driven by dynamical interactions or stellar evolution processes. Additionally, individual stars can participate in multiple ULXs, each with a different companion. This phenomenon is particularly prevalent for IMBHs (see Sect. \ref{sec:imbh}). While our models focus on these long-term changes, it's worth noting that observations have shown both globular cluster and field ULXs can exhibit significant short-term variability on timescales of days to months \citep[e.g.,][]{Maccarone07,Shih10,Brightman2307}.

\begin{table*}
\caption{\label{tab:zams_overview}Progenitor properties: ZAMS progenitors, ULX onset time, and total ULXs for different configurations.}
\centering
\begin{tabular}{lllll}
\hline\hline
 & \multicolumn{2}{c}{First Population} & \multicolumn{2}{c}{Second Population} \\
 & binary & single & binary & single \\
\hline
\multicolumn{5}{c}{\large \nprog} \\
A-D(p) & $148.67\pm92.34\;(36)$ &  & $5.86\pm5.75\;(14)$ &  \\
A-n & $9.29\pm15.49\;(14)$ & $28.75\pm17.07\;(24)$ & $8.00\pm13.75\;(9)$ & $5.55\pm8.08\;(11)$ \\
D-A(p) & $2.22\pm2.28\;(9)$ &  &  &  \\
D-n & $23.31\pm24.33\;(13)$ & $27.96\pm18.79\;(24)$ & $12.33\pm17.78\;(15)$ & $8.18\pm8.44\;(11)$ \\
n-A & $6.73\pm9.08\;(11)$ &  & $4.57\pm6.11\;(7)$ &  \\
n-D & $12.25\pm7.31\;(12)$ &  & $6.54\pm4.82\;(13)$ &  \\
\\
\multicolumn{5}{c}{\large \tulxstart} \\
A-D(p) & $6.21\pm0.44\;(36)$ &  & $297.78\pm209.46\;(14)$ &  \\
A-n & $3275.22\pm3442.16\;(14)$ & $345.72\pm1209.67\;(24)$ & $1726.87\pm3277.15\;(9)$ & $1922.14\pm3139.30\;(11)$ \\
D-A(p) & $6924.70\pm5695.35\;(9)$ &  &  &  \\
D-n & $3752.29\pm2056.81\;(13)$ & $334.86\pm1242.41\;(24)$ & $2074.99\pm1919.42\;(15)$ & $1599.31\pm2702.66\;(11)$ \\
n-A & $3291.82\pm3514.45\;(11)$ &  & $1851.74\pm2567.34\;(7)$ &  \\
n-D & $3626.39\pm2618.49\;(12)$ &  & $1596.63\pm2707.93\;(13)$ &  \\
\\
\multicolumn{5}{c}{\large \enulxtotal} \\
A-D(p) & $88.39\pm162.02\;(36)$ &  & $0.32\pm0.37\;(14)$ &  \\
A-n & $56.30\pm82.00\;(14)$ & $77.75\pm147.37\;(24)$ & $55.16\pm82.91\;(9)$ & $7.79\pm15.10\;(11)$ \\
D-A(p) & $42.20\pm123.25\;(9)$ &  &  &  \\
D-n & $68.97\pm132.00\;(13)$ & $49.56\pm94.57\;(24)$ & $22.28\pm28.24\;(15)$ & $26.22\pm62.54\;(11)$ \\
n-A & $49.22\pm68.55\;(11)$ &  & $2.06\pm1.91\;(7)$ &  \\
n-D & $91.36\pm174.29\;(12)$ &  & $3.01\pm2.27\;(13)$ &  \\
\hline
\end{tabular}
\tablefoot{
Number of progenitors on ZAMS (\nprog), median of the time when the ULX phase commences (\tulxstart), and the total (i.e., through the entire history) expected number of ULXs (\enulxtotal). Number in parenthesis inform about the number of simulations with non-zero \enulx\ of specific configuration. Results are present for different configurations of progenitors: A stands for the progenitor of an accretor, whereas D for the progenitor of donor, n stands for stars which do not become part of a ULX, or is a placeholder in case of single stars. p stands for pristine and means that there were no exchanges in a binary before ULX formation. First symbol represents the primary (i.e., more massive star on zero-age main sequence, or ZAMS), whereas the second letter the secondary. Results are divided into First and Second Populations and also between binaries and single stars. Values represent the mean and one sigma error for simulations in which the specific ULX where present.
}
\end{table*}

Table \ref{tab:zams_overview} provides a comprehensive overview of ULX progenitors on zero-age main sequence (ZAMS), that is stars that later became components of ULXs. The most common progenitors are those of pristine ULXs, that is binaries that do not undergo exchanges or disruptions prior to the ULX phase, although still they can dynamically interact and change their properties. In these systems, the primary typically becomes the accretor and the secondary the donor (A-D(p)). These progenitors evolve rapidly, initiating the ULX phase after approximately $\tulxstart\approx6\myr$, in stark contrast to other progenitors which generally begin after $\tulxstart\gtrsim3\gyr$. Pristine ULXs are observed in all simulations.

Interestingly, some pristine ULXs form where the secondary, less massive on ZAMS, becomes the accretor (D-A(p)). These progenitors are considerably rarer (about two orders of magnitude less common than A-D(p)) and typically form ULXs only in later evolutionary phases ($\sim7\gyr$).

Non-pristine ULXs can originate from both binary and single star progenitors on ZAMS. However, stars born in binaries more frequently become components of ULXs, even in non-pristine cases. These systems invariably undergo significant interactions, such as binary formation, exchanges or disruptions, before forming a ULX. While some of these ULXs can form relatively early ($\sim100\myr$), the majority emerge in later evolutionary phases ($>3\gyr$). In non-pristine ULXs, both the primary and secondary can act as either the accretor or donor with comparable probability. Our simulations did not reveal any non-pristine ULXs where both the primary and secondary are parts of ULXs (i.e., "A-D" or "D-A" type), though this is likely due to limited statistics rather than a physical constraint.

\subsection{ULX properties}\label{sec:ulx_properties}

In this section we discuss the properties of in-cluster ULXs, that is these which reside inside the cluster. For a discussion of ULXs among escapers see Section \ref{sec:escapers}.

\begin{table*}
\tiny
\caption{\label{tab:ulx_groups}Properties of ULXs by group}
\centering
\begin{tabular}{llllllllll}
\hline\hline
 group & \kacc & \kdon & \macc & \mdon & \sep & \ecc & $\log_{10}\lxmax$ & \dtulx & $\tphysmin$ \\
\hline
\gbhms & BH & MS & $27.64^{+18.02}_{-19.43}$ & $45.72^{+81.05}_{-42.83}$ & $40.05^{+104.24}_{-32.82}$ & $0.00^{+0.58}_{-0.00}$ & $40.36^{+0.38}_{-1.11}$ & $0.41^{+4.71}_{-0.41}$ & $6.16^{+380.87}_{-2.86}$ \\
\gbhhg & BH & HG & $22.70^{+22.73}_{-14.22}$ & $56.64^{+46.70}_{-51.47}$ & $119.21^{+905.18}_{-88.52}$ & $0.00^{+0.45}_{-0.00}$ & $40.35^{+0.32}_{-0.84}$ & $0.13^{+1.35}_{-0.13}$ & $5.63^{+131.94}_{-2.74}$ \\
\gbhcheb & BH & CHeB & $16.88^{+29.46}_{-8.38}$ & $32.52^{+43.80}_{-18.12}$ & $233.71^{+2887.84}_{-179.57}$ & $0.03^{+0.36}_{-0.03}$ & $40.18^{+0.38}_{-0.60}$ & $0.10^{+0.34}_{-0.09}$ & $7.52^{+9.56}_{-4.29}$ \\
\gimbh & BH & * & $9351.31^{+4162.72}_{-8738.78}$ & $0.96^{+168.06}_{-0.73}$ & $28.54^{+3054.93}_{-28.26}$ & $0.56^{+0.39}_{-0.56}$ & $40.84^{+2.29}_{-1.76}$ & $0.00^{+3.06}_{-0.00}$ & $2207.25^{+11595.73}_{-2195.15}$ \\
\gns & NS & ** & $1.27^{+1.17}_{-0.16}$ & $0.78^{+4.51}_{-0.51}$ & $0.45^{+51.06}_{-0.43}$ & $0.00^{+0.34}_{-0.00}$ & $39.22^{+0.52}_{-0.21}$ & $0.05^{+27.48}_{-0.05}$ & $331.03^{+12959.06}_{-258.73}$ \\
other & NS or BH & *** & $22.13^{+22.42}_{-20.84}$ & $3.22^{+32.80}_{-2.89}$ & $489.01^{+3081.72}_{-488.99}$ & $0.00^{+0.64}_{-0.00}$ & $39.95^{+0.87}_{-0.85}$ & $0.06^{+32.44}_{-0.06}$ & $143.70^{+13216.21}_{-138.87}$ \\
\hline
\end{tabular}
\tablefoot{
Median properties and $95\%$ confidence intervals for ULX groups identified in this study. Column definitions: \kacc\ - accretor type (BH - black hole, NS - neutron star); \kdon\ - donor type; \macc\ - accretor mass (\msun); \mdon\ - donor mass (\msun); \sep\ - orbital separation (\rsun); \ecc\ - eccentricity; $\lxmax$ - maximum X-ray luminosity (\ergs); \dtulx\ - ULX lifetime (\myr); $\tphysmin$ - ULX phase onset age (\myr). Donor types: MS - main sequence, HG - Hertzsprung gap, RG - red giant, CHeB - core helium burning. For groups with mixed donor types: * HG: $47\%$, MS: $41\%$, RG: $6\%$, CHeB: $3\%$; ** RG: $83\%$, HG: $7\%$, MS: $6\%$; *** RG: $75\%$, CHeB: $21\%$; other companion types represent less then $5\%$ of all companions.
}
\end{table*}

Table \ref{tab:ulx_groups} summarizes the key properties of ULXs in our sample, categorized into different groups based on the types of accretors and donors. The majority of ULXs in our sample are powered by BH accreting from various companion types. BH accretors paired with MS donors (\gbhms) show median masses of $27.64\msun$, while those with HG and CHeB companions (\gbhhg\ and \gbhcheb, respectively) have slightly lower median masses ($22.70\msun$ and $16.88\msun$, respectively). The maximum X-ray luminosities for these systems (median $\log_{10}\lxmax \approx 40.18$ -- $40.36\ergs$) are well above the ULX defining limit of $10^{39}\ergs$. The observed sample of GCULXs, with the brightest having $\lx\approx4\times10^{39}\ergs$ \citep{Dage1905}, appears inconsistent with such high luminosities. However, it is important to note that the simulated ULXs exhibiting these extreme luminosities primarily form in very young clusters ($\tphys\lesssim10\myr$), whereas the observed GCULXs are found in older globular clusters. Our simulations predict $<0.1$ \enulx\ in older clusters ($8$--$13\gyr$) similar to Milky Way globular clusters, and comparable limit for intermediate-age clusters ($2$--$8\gyr$). Additionally, high-luminosity ULXs are typically short-lived, further reducing their detection probability.

Within ULXs, NS accretors (\gns) have median masses around $1.27\msun$. These systems typically have lower X-ray luminosities ($\log_{10}\lxmax \approx 39.22$) compared to their BH counterparts, but still clearly exceed the Eddington limit for a typical NS.

The orbital characteristics of ULXs vary significantly across different groups. \gbhms\ systems have median separations of $40.05\rsun$, while \gns\ systems show much tighter orbits (median $0.45\rsun$). Most ULXs in our sample have low eccentricities, suggesting that tidal forces or a common envelope phase have circularized their orbits before the onset of the ULX phase.

Duty cycles vary widely among ULX groups, with \gbhms\ systems showing the highest median value ($0.41$), while other groups have much lower values. This variability in duty cycles may explain the transient nature observed in many ULXs and has implications for their detectability \citep[cf.][]{Wiktorowicz1709}.

A notable subset of our sample consists of IMBH-powered systems (\gimbh). These ULXs are characterized by extremely high accretor masses (median $9351\msun$) and the highest X-ray luminosities in our sample ($\log_{10}\lxmax \approx 40.84$). Such luminous systems, called extreme-ULXs \citep{Wiktorowicz1904}, are one of the main observational candidates for IMBHs and can provide insights into their formation and properties.

The diversity in donor types across ULX groups highlights the various evolutionary pathways that can lead to ULX formation. Donors of types MS, HG, and RG are all well-represented, suggesting that ULXs can form and persist across a wide range of stellar evolutionary stages.

Huge majority are ULXs with BH accretors (typically $>80\%$) formed early in cluster evolution.

Table \ref{tab:ulx_group_fracions} shows a modest anticorrelation between the relative size of the \gimbh\ and BH ULXs (Kendall\footnote{Kendall's $\tau$ is preferable to Pearson's $r$ when working with small samples and non-normal distributions, while still providing a similarly interpretable measure of association between $-1$ and $+1$.}'s $\tau=-0.49, -0.38, -0.37$, for \gbhms, \gbhhg, and \gbhcheb, respectively). The presence of IMBH results in a continuous formation of IMBH powered ULX though the cluster history which, being an independent source to other BH ULX groups present mostly in the early cluster evolution phases, lowers their fractional input to the ULX population. IMBH quickly removes BHs and possibly NSs from the system \citep[e.g.,][]{Hong2011}.

\subsection{IMBH ULXs}\label{sec:imbh}

In this study, we define IMBHs as those with masses exceeding $500\msun$. IMBH ULXs were observed in 6 simulations, all featuring multiple stellar populations (npop2) and non-tidally filling initial conditions (nTF). When present, IMBHs contribute significantly to the ULX population, accounting for 16-44\% of ULXs over a Hubble time (Table \ref{tab:ulx_group_fracions}).

IMBHs form rapidly, within several Myr after the ZAMS, classifying them as "fast IMBHs" \citep{Giersz1512}. In simulations with delayed second populations, IMBH formation occurs 100+ Myr after the initial population. Notably, IMBH ULXs form exclusively in non-tidally filling (nTF) simulations, strenghtening the preference for denser stellar environments.

Fisher's exact tests revealed no significant correlations between simulation parameters and IMBH formation at the 90\% confidence level. However, when considering only simulations with multiple populations (pop=2) and active dynamics, a statistically significant relationship (95\% confidence) emerged between the \texttt{filling} parameter and IMBH ULX presence, favoring non-tidally filling clusters.

\begin{table*}
\small
\caption{\label{tab:imbh}IMBH ULX properties}
\centering
\begin{tabular}{lrrrrrl}
\hline\hline
label & \tphysmin& \tphysmax & $\text{M}_\mathrm{min}$ & $\text{M}_\mathrm{max}$ & duty-cycle [\%] & companions \\
\hline
npop2-cpop05-fb10-nTF & $0.008$ & $13.3$ & $3.8$ & $13.0$ & $0.06$ & MS: 22, HeWD: 4, COWD: 3,\\
 & & & & & & TPAGB: 2, CHeB: 2 \\
npop2-cpop05-fb10-nTF-NF & $0.1$ & $14.9$ & $0.5$ & $13.5$ & $0.02$ & MS: 41, COWD: 5, RG: 3,\\
& & & & & & HeWD: 2, CHeB: 1, EAGB: 1,\\
& & & & & & TPAGB: 1, ONeWD: 1 \\
npop2-cpop05-fb95-nTF & $0.01$ & $13.8$ & $5.0$ & $13.5$ & $0.87$ & MS: 29, COWD: 13, HG: 1,\\
& & & & & & HeWD: 1, RG: 1 \\
npop2-cpop05-fb95-nTF-NF & $0.1$ & $14.8$ & $0.6$ & $12.9$ & $0.22$ & MS: 89, COWD: 12, TPAGB: 2,\\
& & & & & & HeWD: 2, CHeB: 2, HeGB: 1,\\
& & & & & & ONeWD: 1, HG: 1 \\
npop2-cpop2-fb10-nTF & $0.009$ & $14.2$ & $0.6$ & $4.5$ & $0.14$ & MS: 32, COWD: 5, HeWD: 2,\\
& & & & & & RG: 1, TPAGB: 1, ONeWD: 1 \\
npop2-cpop2-fb95-nTF & $0.009$ & $14.5$ & $2.1$ & $6.8$ & $0.43$ & MS: 43, COWD: 11, HG: 3,\\
& & & & & & HeWD: 3, TPAGB: 2, RG: 2 \\
\hline
\end{tabular}
\tablefoot{
Properties of $6$ intermediate-mass black holes (IMBHs) ULXs observed in our simulations. Each IMBH forms multiple, short-lived ULX phases with different companions. Column definitions: label - simulation model name; \tphysmin\ - minimum ULX phase onset age (\gyr); \tphysmax\ - maximum age at which ULX phases occur (\gyr); $\text{M}_\mathrm{min}$ - minimum IMBH mass during ULX phases ($1000\;\times\msun$); $\text{M}_\mathrm{max}$ - maximum IMBH mass during ULX phases ($1000\;\times\msun$); duty-cycle - percentage of total IMBH lifetime spent in ULX phases; companions - types and numbers of donor stars (MS - main sequence, HG - Hertzsprung gap, RG - red giant, CHeB - core helium burning, EAGB - early asymptotic giant branch, TPAGB - thermally pulsating asymptotic giant branch, HeGB - helium giant branch, HeWD - helium white dwarf, COWD - carbon-oxygen white dwarf, ONeWD - oxygen-neon white dwarf).
}
\end{table*}

Table \ref{tab:imbh} presents key properties of the six IMBHs observed in our simulations. These IMBHs display a substantial mass range, from as low as $\sim500\msun$ to as high as $\sim13,500\msun$, demonstrating their significant growth potential within globular cluster environments. IMBH ULXs typically exhibit intermittent ULX phases (duty-cycle between $0.02$ -- $0.87\%$) with various donors, interspersed with periods of non-ULX mass accretion or mergers. Donors are mostly MS stars, with white dwarfs (particularly COWD) being the second most common companion type.

IMBH ULXs formed exclusively in simulations with two stellar populations and non-tidally filling initial conditions. The centrally concentrated second population facilitates stellar interactions and mergers, while nTF clusters retain more stars, particularly BHs, in central regions, extending the IMBH formation window.

IMBH ULXs typically exist throughout the entire GC life, with first ULX phases appearing as early as $\sim8\myr$ in some models and continuing until the end of our simulations at $\sim14\gyr$, indicating that IMBH ULXs can form very late in GC evolution.

\subsection{ULX progenitors}

The initial properties of binary systems that evolve into ULXs provide crucial insights into their formation channels and evolutionary pathways. Our results indicate that the primary distinguishing factor between ULX progenitors and the general stellar population is their higher initial mass. No other significant constraints on ZAMS properties were identified for ULX progenitors.

\begin{table*}
\caption{\label{tab:progenitor_groups}Initial properties of ULX progenitors by group}
\centering
\begin{tabular}{lllrrrrrlll}
\hline\hline
 group & * & \mzams & \fesc & \fbin & mode & \fprim & \fpris & \mcomp & \sep & \ecc \\
\hline
\gbhms & A & $47.00^{+73.39}_{-46.10}$ & $0.07$ & $0.91$ & A-D(p) (85\%) & $0.98$ & $0.94$ & $20.89^{+64.91}_{-12.95}$ & $71.01^{+3187.79}_{-39.51}$ & $0.19^{+0.67}_{-0.19}$ \\
 & D & $18.48^{+59.00}_{-17.59}$ & $0.07$ & $0.91$ & A-D(p) (83\%) & $0.07$ & $0.91$ & $48.97^{+71.42}_{-47.11}$ & $67.00^{+3191.81}_{-42.35}$ & $0.18^{+0.68}_{-0.18}$ \\
\\
\gbhhg & A & $48.13^{+97.30}_{-40.02}$ & $0.03$ & $0.96$ & A-D(p) (95\%) & $0.99$ & $0.98$ & $33.22^{+78.41}_{-24.70}$ & $115.21^{+2330.09}_{-79.31}$ & $0.13^{+0.60}_{-0.13}$ \\
 & D & $32.29^{+79.34}_{-30.81}$ & $0.03$ & $0.97$ & A-D(p) (94\%) & $0.01$ & $0.98$ & $48.59^{+96.84}_{-27.64}$ & $114.91^{+2149.83}_{-79.93}$ & $0.14^{+0.59}_{-0.14}$ \\
\\
\gbhcheb & A & $37.02^{+76.30}_{-36.45}$ & $0.11$ & $0.89$ & A-D(p) (88\%) & $1.00$ & $0.99$ & $23.69^{+48.14}_{-12.46}$ & $398.98^{+3919.72}_{-350.69}$ & $0.21^{+0.60}_{-0.21}$ \\
 & D & $21.93^{+49.90}_{-21.37}$ & $0.11$ & $0.89$ & A-D(p) (88\%) & $0.01$ & $0.99$ & $37.98^{+75.81}_{-16.70}$ & $398.98^{+3897.78}_{-352.62}$ & $0.21^{+0.60}_{-0.21}$ \\
\\
\gimbh & A & $13.54^{+2.57}_{-3.50}$ & $0.00$ & $0.67$ & A-n (100\%) & $1.00$ & $0.00$ & $9.67^{+2.20}_{-0.85}$ & $75.45^{+1.96}_{-39.57}$ & $0.06^{+0.07}_{-0.05}$ \\
 & D & $1.19^{+14.35}_{-1.05}$ & $0.00$ & $0.88$ & D-n (61\%) & $0.56$ & $0.00$ & $0.90^{+11.26}_{-0.72}$ & $36.70^{+276433.30}_{-30.94}$ & $0.34^{+0.58}_{-0.33}$ \\
\\
\gns & A & $5.98^{+12.30}_{-5.87}$ & $0.39$ & $0.68$ & A-D(p) (49\%) & $0.90$ & $0.74$ & $3.04^{+8.86}_{-2.94}$ & $1410.90^{+600401.60}_{-1393.38}$ & $0.32^{+0.58}_{-0.32}$ \\
 & D & $2.07^{+8.32}_{-1.98}$ & $0.39$ & $0.70$ & A-D(p) (48\%) & $0.27$ & $0.71$ & $5.46^{+11.13}_{-5.33}$ & $1410.90^{+724659.10}_{-1400.20}$ & $0.34^{+0.58}_{-0.34}$ \\
\\
other & A & $24.70^{+46.22}_{-24.56}$ & $0.15$ & $0.80$ & A-D(p) (57\%) & $0.89$ & $0.73$ & $10.44^{+40.18}_{-10.24}$ & $1849.40^{+226115.60}_{-1824.04}$ & $0.22^{+0.70}_{-0.22}$ \\
 & D & $4.73^{+31.64}_{-4.61}$ & $0.14$ & $0.80$ & A-D(p) (52\%) & $0.28$ & $0.66$ & $22.43^{+46.86}_{-22.31}$ & $1409.40^{+338400.60}_{-1396.45}$ & $0.31^{+0.59}_{-0.31}$ \\
\hline
\end{tabular}
\tablefoot{
Initial properties of ULX progenitors, including median values with $95\%$ confidence intervals or averages. Groups are defined in Table \ref{tab:ulx_groups} and accompaning text. * denotes component (A - accretor, D - donor). \mzams: ZAMS mass (\msun); esc: fraction of ULXs formed from escapers; bin: binary fraction among progenitors; mode: dominant formation mode (see Table \ref{tab:zams_overview}); \fprim: fraction of progenitors formed from primary stars; \fpris: fraction of pristine ULXs formed; \mcomp: companion ZAMS mass (\msun); \sep: ZAMS separation (\rsun); \ecc: ZAMS eccentricity. Mode, \fprim, \fpris, \mcomp, \sep, and \ecc\ values are calculated only for binary progenitors.
}
\end{table*}

Table \ref{tab:progenitor_groups} summarizes the ZAMS properties of ULX progenitors, categorized by ULX group and component role (accretor or donor). Accretor progenitors typically have higher ZAMS masses (\mzams) than their companions (\mcomp), consistent with being predominantly formed from primary stars ($\fprim \geq 0.89$).

The majority of ULX progenitors originate in binary systems ($\fbin \geq 0.67$), with most maintaining their original pairing to form the ULX ($\fpris \geq 0.66$, except for the \gimbh\ group). For accretors in the \gimbh\ and \gns\ groups, $\fbin \approx 0.66$, indicating no preference between single and binary star origins.

Initial separations (\sep) exhibit a highly skewed distribution, with \gns\ group progenitors typically having much wider separations than other groups. Values for the \gimbh\ group are not particularly informative as these stars do not become components of ULXs without undergoing strong dynamical interactions. All groups show a preference for lower initial eccentricities ($\med{\ecc}\lesssim 0.3$).

The \gimbh\ group shows distinct characteristics compared to other groups. It has the lowest ZAMS masses for both accretors and donors, a unique dominant formation modes (A-n for accretors and D-n for donors), and the lowest \fbin. The \gbhms\ and \gbhhg\ groups exhibit similar patterns, with high ZAMS masses and a strong preference for the A-D(p) formation mode. The \gns\ group has the highest fraction of ULXs formed from escapers ($\fesc = 0.39$).

\subsection{ULX descendants}

\begin{table}
\caption{\label{tab:descendants}ULX descendants}
\centering
\begin{tabular}{lrrrrr}
\hline\hline
 & D & E & X & M & O\\
\hline
npop1-fb10-TF &  & $67$ &  & $14$ &  \\
npop1-fb10-nTF & $4$ & $22$ & $46$ & $25$ & $4$ \\
npop1-fb95-TF &  & $222$ &  & $36$ & $1$ \\
npop1-fb95-nTF & $18$ & $69$ & $152$ & $70$ & $19$ \\
npop2-cpop05-fb10-TF & $2$ & $35$ & $11$ & $13$ &  \\
npop2-cpop05-fb10-TF-NF & $2$ & $25$ & $21$ & $14$ & $2$ \\
npop2-cpop05-fb10-nTF & $18$ & $20$ & $10$ & $42$ & $1$ \\
npop2-cpop05-fb10-nTF-NF & $23$ & $12$ & $10$ & $65$ & $1$ \\
npop2-cpop05-fb95-TF & $4$ & $117$ & $30$ & $37$ & $3$ \\
npop2-cpop05-fb95-TF-NF & $13$ & $90$ & $49$ & $37$ & $9$ \\
npop2-cpop05-fb95-nTF & $55$ & $72$ & $38$ & $81$ & $3$ \\
npop2-cpop05-fb95-nTF-NF & $66$ & $54$ & $41$ & $136$ & $8$ \\
npop2-cpop2-fb10-TF &  & $37$ & $2$ & $12$ & $3$ \\
npop2-cpop2-fb10-TF-NF & $3$ & $24$ & $15$ & $14$ & $1$ \\
npop2-cpop2-fb10-nTF & $17$ & $14$ & $16$ & $55$ & $2$ \\
npop2-cpop2-fb10-nTF-NF & $9$ & $27$ & $60$ & $67$ & $12$ \\
npop2-cpop2-fb95-TF & $2$ & $115$ & $17$ & $34$ & $9$ \\
npop2-cpop2-fb95-TF-NF & $4$ & $88$ & $42$ & $36$ & $13$ \\
npop2-cpop2-fb95-nTF & $73$ &  & $41$ & $105$ & $66$ \\
npop2-cpop2-fb95-nTF-NF & $17$ & $60$ & $124$ & $130$ & $45$ \\
\hline
Approximate Average & $19$ & $62$ & $40$ & $51$ & $11$ \\
\hline
\end{tabular}
\tablefoot{
Distribution of ULX descendant fates for various simulation configurations. Abbreviations: D - disruption; E - escape; X - exchange; M - merger; O - other. Only simulations with dynamical interactions are included. The last row represents approximate averages across all simulations. See text for discussion and details.
}
\end{table}

The fate of ULX systems after their active phase provides valuable insights into their evolutionary pathways and potential contributions to other astrophysical phenomena. 

Table \ref{tab:descendants} summarizes the various outcomes for ULX systems across different simulation configurations. The table shows, that ULX systems can undergo several fates: disruption, escape from the cluster, exchange interactions, mergers, or other outcomes. The distribution of these fates appears largely stochastic across different simulation parameters. Systems categorized as "other" typically experience no significant events for the remainder of their evolution and are predominantly NS/BH + WD or double compact objects at the end of simulation.

Merger events typically occur $0.79^{+0.74}_{-0.53}\myr$ after the ULX phase. Double compact objects form in $\sim63\%$ of cases, with $\sim34\%$ of them being ejected from the cluster before merger or the end of simulation. These double compact objects that remain in the cluster may contribute to the population of gravitational wave sources with merger rate of $\sim11\%$. 

Disruptions may result from dynamical interactions or binary evolution. In exchanges, at least one of the components continues evolution as part of some other binary. Our results suggest, that in initially more centrally concentrated environment (nTF) there are more disruptions and exchanges after the ULX phases, then in other simulations.

\subsection{ULXs among escapers}\label{sec:escapers}

Cluster-formed binaries are frequently ejected, suggesting that some field ULXs may originate from GCs. These ULX progenitors become unbound from their parent clusters and initiate the ULX phase while residing in the galactic field, evolving as isolated binaries.

Figure \ref{fig:n_ulx_grid} illustrates the expected numbers and ages of ULXs among escapers. Generally, escaper ULXs coexist in similar numbers and at comparable evolutionary times as their in-cluster counterparts, except the very early phase of cluster evolution ($\lesssim300\myr$). Table \ref{tab:escapers} presents a detailed comparison of properties between in-cluster and escaper ULXs.

\begin{table*}
\caption{\label{tab:escapers}Properties of escaper and in-cluster ULXs}
\centering
\begin{tabular}{lllll}
\hline\hline
 & \multicolumn{2}{c}{escaper} & \multicolumn{2}{c}{in-cluster} \\
 & NS & BH & NS & BH \\
\hline
    count & $12.50^{+25.12}_{-10.55}$ & $19.00^{+20.35}_{-15.05}$ & $7.00^{+89.20}_{-6.00}$ & $142.50^{+158.18}_{-88.50}$ \\
    Typ. companion & HeMS (70\%) & CHeB (80\%) & HG (74\%) & HG (55\%) \\
    $\med{\macc}$ & $1.32^{+0.52}_{-0.18}$ & $16.58^{+16.05}_{-1.98}$ & $1.26^{+0.34}_{-0.09}$ & $22.51^{+147.80}_{-4.10}$ \\
    $\med{\mdon}$ & $0.54^{+1.87}_{-0.06}$ & $19.00^{+15.82}_{-15.04}$ & $1.80^{+6.95}_{-1.41}$ & $40.55^{+7.59}_{-23.52}$ \\
    $\med{\sep}$ & $0.43^{+14.08}_{-0.40}$ & $126.33^{+89.54}_{-88.43}$ & $16.55^{+18.52}_{-16.51}$ & $123.62^{+76.99}_{-76.17}$ \\
    $\med{\ecc}$ & $0.00^{+0.11}_{-0.00}$ & $0.09^{+0.26}_{-0.09}$ & $0.01^{+0.21}_{-0.01}$ & $0.00^{+0.03}_{-0.00}$ \\
    $\med{\log_{10}\lxmax}$ & $39.22^{+0.24}_{-0.10}$ & $40.05^{+0.09}_{-0.06}$ & $39.20^{+0.17}_{-0.19}$ & $40.28^{+0.11}_{-0.14}$ \\
    $\med{\dtulx}$ & $0.12^{+0.08}_{-0.11}$ & $0.12^{+0.18}_{-0.08}$ & $0.05^{+10.55}_{-0.05}$ & $0.11^{+0.02}_{-0.07}$ \\
    $t_\mathrm{ULX, min}$ & $115.30^{+120.96}_{-73.08}$ & $3.49^{+4.99}_{-0.60}$ & $119.45^{+1512.99}_{-106.11}$ & $2.85^{+0.48}_{-0.13}$ \\
    $t_\mathrm{ULX, max}$ & $5767.74^{+9100.42}_{-5558.67}$ & $1994.71^{+11988.23}_{-1953.63}$ & $11682.46^{+3038.02}_{-11616.14}$ & $11507.08^{+3391.92}_{-11440.52}$ \\
\hline
\end{tabular}
\tablefoot{
Comparison of properties between ULXs in-cluster and those among escapers. Values represent the median with errors indicating the 10th and 90th percentiles of the all simulations with dynamics included. Most of the values represent the median values for the simulation ($\med{}$). \macc/\mdon\ - accretor/donor mass [\msun], \sep\ - orbital separation [\rsun], \ecc\ - eccentricity, \lxmax\ - maximum X-ray luminosity [\ergs], $t_\mathrm{ULX, min}$/$t_\mathrm{ULX, max}$\ - minimum/maximum age of ULX phase onset [\myr].
}
\end{table*}

Escaper ULXs exhibit several distinct characteristics compared to their in-cluster counterparts. They tend to have lower-mass accretors and donors, with median masses of $13.00$ and $6.05\msun$, respectively, compared to $22.24$ and $39.34\msun$ for in-cluster ULXs. Escaper ULXs also have a significantly higher fraction of NS accretors ($40\%$ vs. $4\%$) and tighter orbits (median $57.42$ vs. $119.93\rsun$). This orbital difference largely stems from dynamical hardening processes that occur prior to ejection, whereby close encounters cause binaries to become more tightly bound \citep{Heggie7512}. Additionally, the ejection mechanisms preferentially remove harder binaries with sufficient kinetic energy to escape the cluster potential. The median maximum X-ray luminosity of escaper ULXs ($\log_{10}\lxmax = 39.89\ergs$) is slightly lower than that of in-cluster ULXs ($\log_{10}\lxmax = 40.26\ergs$).

Notably, escaper ULXs are less numerous, with a median count of $31$ compared to $176$ for in-cluster ULXs. They also typically have more evolved companion stars, with $65\%$ having CHeB donors compared to $55\%$ HG donors for in-cluster ULXs. The duty cycles and ULX phase onset times are similar for both populations.

The ejection processes preferentially remove less massive systems, explaining the lower masses observed in escaper ULXs. This mass-dependent retention occurs because more massive systems require stronger dynamical encounters to achieve escape velocity from the cluster potential. The higher fraction of NS accretors among escapers could be due to the higher natal kicks and retention of more massive BHs within the cluster potential.

These findings have important implications for understanding the origin and evolution of field ULXs. A significant fraction of observed field ULXs may have originated in GCs, which could explain some of the observed diversity in ULX populations. While our current analysis captures the properties of escaper ULXs at the moment of escape and during the ULX phase, tracking their complete galactic trajectories would require additional post-processing to integrate their orbits through the galactic potential \citep[e.g.,][]{Cabrera2308}. Furthermore, the differences between escaper and in-cluster ULXs provide valuable insights into how environment and dynamical history can influence the properties of these extreme systems.

In conclusion, our simulations reveal that escaper ULXs form a significant and distinct population with properties that differ from their in-cluster counterparts. These differences provide valuable insights into the formation and evolution of ULXs in various environments and highlight the need for comprehensive observational surveys to fully characterize the ULX population across different galactic settings.

\section{Discussion}

The results presented in this study are specific to the simulations employed and may not capture general trends applicable to all GC environments. The parameter choices were carefully tailored to conditions characteristic of GCs, informed by insights from analogous studies.

A key limitation in the broader study of ULXs in GCs is the dearth of confirmed observations within the Milky Way and its immediate vicinity. The brightest Milky Way X-ray binaries fall well below the ULX luminosity threshold \citep{Fortin2303,Fortin2404}. The substantial distances involved often complicate the detection and characterization of low-luminosity donor stars, which are believed to comprise a significant fraction of ULX donors \citep[e.g.,][]{Wiktorowicz1709}.

Furthermore, a critical challenge for this study lies in the difficulty of unambiguously differentiating GC-associated ULXs from those in the field environment. Observational uncertainties frequently obscure the distinction, as localization on the plane of a GC does not inherently confirm membership within the cluster. Determining the precise distance to the source is often fraught with uncertainties, leaving room for the possibility that the ULX is a foreground or background source, merely coincident with the GC in projection. This ambiguity highlights the importance of complementary approaches, such as proper motion studies or radial velocity measurements, to confirm GC membership conclusively.

\subsection{Comparison with field populations}

We find a high fraction of NS accretors among escapers ($40\%$) compared to in-cluster ULXs ($4\%$). This is similar to the predictions of \citet{Wiktorowicz1904}, who found that NS ULXs outnumber BH ULXs in regions with constant star formation and solar metallicity for ages above $\sim1\gyr$. 

The properties of our escaper NS ULXs show some similarities with typical field NS ULXs described by \citet{Wiktorowicz1709}. They found that field NS ULXs typically have $\sim1.3 \msun$ NS accretors and $\sim1.0 \msun$ red giant donors. This is comparable to our escaper ULXs with NS accretors, which have low-mass $\lesssim2\msun$ evolved (HeMS) donors. Our systems are very compact ($\sep\lesssim15\rsun$) which may explain the lower fraction of giant donors.

Our escaper NS ULXs have a median maximum X-ray luminosity of $\log_{10}\lxmax = 39.22\ergs$, with a few BH ULXs exceeding $10^{40}\ergs$. This stays in contrast to the field populations, where a significant fraction of more luminous sources is also expected \citep[e.g.,][]{Wiktorowicz1509}. In the case of GCs, we find that such extreme ULXs are more likely to be retained within the cluster.

The formation pathways of escaper ULXs may differ from those in the field. For instance, \citet{Fragos1503} studied the formation of NS ULXs like M82 X-2, finding that systems with $8$ -- $10\msun$ donor stars could produce such ULXs. In contrast, our escaper NS ULXs have lower donor masses ($\lesssim2\msun$), suggesting a different formation pathway.

The temporal distribution of ULXs in our simulations differs from that observed in field populations. \citet{Kuranov2112} found that the maximum number of ULXs ($\sim10$ for a star formation rate of $10\msun/yr$) is reached $\sim1\gyr$ after the beginning of star formation in field populations. Our escaper ULXs show a different temporal distribution with a more uniform appearance of escaper ULXs through the GC lifespan.

It is worth noting that even field populations are not entirely free from dynamical influences. \citet{Klencki1708} showed that wide binaries in the field are affected by multiple weak interactions (fly-bys) which can impact binary evolution. This suggests that the distinction between cluster-formed and field-formed ULXs may not be as clear-cut as previously thought, and that dynamical effects could play a role in shaping ULX populations across various environments.

\subsection{Counterparts}

Identifying ULX counterparts is crucial yet challenging \citep{Heida1606,Wiktorowicz2109}. While field ULXs face issues of faintness and crowded environments \citep{Lopez2009,Allak2209,Heida1902}, extragalactic GC ULXs present unique challenges due to their unresolved stellar populations. However, GCs' well-understood stellar demographics suggest these ULXs are likely LMXBs. Rare cases allow indirect donor identification, such as oxygen emission indicating a WD donor in a NGC 4472 GC ULX \citep{Steele1404}. Accurate astrometry remains essential for all ULX studies \citep{Allak2209}, and most counterparts, in both field and GCs, remain tentative due to observational biases \citep{Heida1902,Wiktorowicz2109}.

\subsection{Expectations for Antennae galaxy}

The correlation between X-ray source positions and stellar clusters has been previously noted in starburst galaxies \citep[e.g.,][]{Kaaret0402}. \citet{Poutanen1306}'s study of the Antennae galaxies reveals a significant association between bright X-ray sources and stellar clusters, with most X-ray sources located outside clusters. This supports the idea that many ULXs are massive X-ray binaries ejected from their birth clusters, rather than IMBHs.

The displacements of ULXs from cluster centers, ranging up to $300$ parsecs, are statistically significant and likely due to ejection of massive binaries. The ejection mechanisms suggested by \citet{Poutanen1306}, including supernova kicks and few-body encounters, are consistent with our results showing higher NS accretor fractions and smaller separations among escaped ULXs.

The distribution of ULXs likely evolves with stellar population age. While the Antennae study focus on young populations, older stellar systems are expected to decrease their X-ray binary populations \citep[e.g.,][]{Fragos0808}, highlighting the importance of considering population age in ULX distribution studies.

\subsection{Future prospects}

Future studies of ULXs in GCs should include more simulations to better estimate statistical effects, errors, and the significance of conclusions. A more consistent survey of the parameter space is needed, as well as the incorporation of evolutionary  parameters (both stellar and binary evolution) beyond dynamical ones.

In the future work we plan to include the effect of beaming in ULXs \citep[e.g.,][]{Lasota2312}. Beaming increases the apparent luminosity of ULXs, but decreases the probability of observation due to possible misalignment \citep{Wiktorowicz1904,Khan2201}. Population synthesis simulations, \citet[e.g.,][]{Wiktorowicz1904} indicate that the majority of NS ULXs are beamed. The beaming factor is dependent on the mass transfer rate, with higher mass transfer rates resulting in stronger beaming \citep{King0902}.

Additionally, the importance of wind Roche-lobe overflow \citep[see e.g.,][and references therein]{Wiktorowicz2109,Zuo2105} for NS ULXs with (super)giant donors should be explored. Wind Roche-lobe overflow can lead to stable mass transfer even with large mass ratios, and can boost mass transfer rates to reach ULX luminosity levels.

\section{Conclusions}

This study presents the first numerical investigation into the formation and evolution of ULXs within GCs, utilizing a subset of simulations to explore parameter dependencies.

Our simulations reveal that dynamical interactions may play a critical role in ULX formation, particularly in environments with high stellar density. Even if the initial binary fraction is low, the interactions can produce many ULXs. On average, we find that approximately $96\%$ of ULXs in our simulations have BH companions and the number can be even higher for very young clusters ($\lesssim300\myr$). Among escaper, ULXs have a much higher fraction of NS accretors ($\sim40\%$).

Our simulations show that the ratio of escaper ULXs to in-cluster ULXs is approximately 1:7, but nearly 2:1 for ULX with NS accretors. This ratio is the highest in tidally filling clusters, even with high initial binary fractions, where rapid ejections can hinder in-cluster ULX formation by ejecting the progenitors.

Our findings suggest that the relative scarcity of ULXs observed in GCs may be attributed to the advanced age of their stellar populations. In contrast, field populations usually have continuous star formation.

The apparent absence of ULXs in Milky Way GCs aligns more closely with our models of initially tidally filling clusters rather than non-tidally filling ones. This distinction is crucial: tidally filling clusters experience significant early mass loss through tidal stripping by their host galaxy's gravitational field, leading to the ejection of potential ULX progenitors before they can evolve into ULXs. In contrast, non-tidally filling clusters maintain higher central densities for most of their life, allowing for enhanced ULX formation through dynamical interactions. The presence of IMBH ULXs exclusively in non-tidally filling simulations further emphasizes the importance of maintaining dense stellar environments for certain ULX formation channels.

Furthermore, our results suggest that field populations may be significantly "polluted" by ULXs ejected from GCs (escapers), which could contribute to the relative underrepresentation of ULXs in GCs compared to field environments. Escapers can have properties very similar to the field populations or form distinct configurations.

Future observational efforts should prioritize the comparison of ULX properties across diverse environments, including GCs and the galactic field. Such studies will be critical for validating our theoretical predictions and advancing our understanding of the diverse formation channels of ULXs.

\begin{acknowledgements}
We are grateful to anonymous referee for their insightful comments which helped to improve the paper. GW, MG, AH, LH were supported by the Polish National Science Center (NCN) through the grant 2021/41/B/ST9/01191
AA acknowledges support for this paper from project No. 2021/43/P/ST9/03167 co-funded by the Polish National Science Center (NCN) and the European Union Framework Programme for Research and Innovation Horizon 2020 under the Marie Skłodowska-Curie grant agreement No. 945339. For the purpose of Open Access, the authors have applied for a CC-BY public copyright license to any Author Accepted Manuscript (AAM) version arising from this submission. 
\end{acknowledgements}

\section*{Data availability}\label{sec:data_availability}
The data underlying this article are available in Zenodo at https://doi.org/10.5281/zenodo.14953837. This includes simulation data on ULX systems, globular cluster properties, and initial binary conditions, along with corresponding header files containing column definitions.

\bibliographystyle{aa}
\bibliography{ms}

\begin{appendix}

\begin{table*}[!h]
\section{Additional tables}  
\small
\caption{\label{tab:simulations}Model parameters and initial properties}
\centering
\begin{tabular}{llllrrrrrrr}
\hline\hline
Label & \npop & \concpop & \fracb & \rgc & \rh & $\smt / 10^6$ & \rc & $\xrc/10^3$ & $\log_{10}(\roc)$ & \threl \\
\hline
npop1-fb10-TF & 1 &  & 10\% & $2.01$ & $11.90$ & $0.69$ & $5.85$ & $89.10$ & $2.64$ & $10,888.00$ \\
npop1-fb10-nTF & 1 &  & 10\% & $2.01$ & $1.00$ & $0.69$ & $0.49$ & $89.10$ & $5.87$ & $265.22$ \\
npop1-fb95-TF & 1 &  & 95\% & $1.96$ & $11.93$ & $0.72$ & $5.81$ & $93.93$ & $2.61$ & $6,386.70$ \\
npop1-fb95-nTF & 1 &  & 95\% & $1.96$ & $1.00$ & $0.72$ & $0.49$ & $93.93$ & $5.84$ & $155.06$ \\
npop2-cpop05-fb10-TF & 2 & 0.05 & 10\% & $2.01$ & $11.69$ & $0.66$ & $0.13$ & $12.71$ & $6.35$ & $6,505.30$ \\
npop2-cpop05-fb10-TF-NF & 2 & 0.05 & 10\% & $2.40$ & $11.81$ & $0.46$ & $6.26$ & $68.94$ & $2.41$ & $9,157.70$ \\
npop2-cpop05-fb10-nTF & 2 & 0.05 & 10\% & $2.01$ & $0.99$ & $0.66$ & $0.01$ & $12.71$ & $9.57$ & $160.33$ \\
npop2-cpop05-fb10-nTF-NF & 2 & 0.05 & 10\% & $2.40$ & $1.00$ & $0.46$ & $0.53$ & $68.94$ & $5.63$ & $225.73$ \\
npop2-cpop05-fb95-TF & 2 & 0.05 & 95\% & $1.96$ & $11.75$ & $0.69$ & $0.12$ & $13.03$ & $6.37$ & $3,790.60$ \\
npop2-cpop05-fb95-TF-NF & 2 & 0.05 & 95\% & $2.35$ & $11.77$ & $0.48$ & $6.18$ & $71.69$ & $2.39$ & $5,342.70$ \\
npop2-cpop05-fb95-nTF & 2 & 0.05 & 95\% & $1.96$ & $1.00$ & $0.69$ & $0.01$ & $13.03$ & $9.59$ & $93.85$ \\
npop2-cpop05-fb95-nTF-NF & 2 & 0.05 & 95\% & $2.35$ & $1.00$ & $0.48$ & $0.53$ & $71.69$ & $5.61$ & $132.31$ \\
npop2-cpop2-fb10-TF & 2 & 0.2 & 10\% & $2.01$ & $11.80$ & $0.66$ & $0.50$ & $12.84$ & $4.55$ & $7,083.50$ \\
npop2-cpop2-fb10-TF-NF & 2 & 0.2 & 10\% & $2.40$ & $11.81$ & $0.46$ & $6.09$ & $64.82$ & $2.40$ & $9,157.10$ \\
npop2-cpop2-fb10-nTF & 2 & 0.2 & 10\% & $2.01$ & $1.00$ & $0.66$ & $0.04$ & $12.84$ & $7.77$ & $174.58$ \\
npop2-cpop2-fb10-nTF-NF & 2 & 0.2 & 10\% & $2.40$ & $1.00$ & $0.46$ & $0.52$ & $64.82$ & $5.61$ & $225.72$ \\
npop2-cpop2-fb95-TF & 2 & 0.2 & 95\% & $1.96$ & $11.77$ & $0.69$ & $0.49$ & $13.12$ & $4.56$ & $4,079.50$ \\
npop2-cpop2-fb95-TF-NF & 2 & 0.2 & 95\% & $2.35$ & $11.77$ & $0.48$ & $6.12$ & $70.32$ & $2.35$ & $5,342.50$ \\
npop2-cpop2-fb95-nTF & 2 & 0.2 & 95\% & $1.96$ & $1.00$ & $0.69$ & $0.04$ & $13.12$ & $7.77$ & $101.01$ \\
npop2-cpop2-fb95-nTF-NF & 2 & 0.2 & 95\% & $2.35$ & $1.00$ & $0.48$ & $0.52$ & $70.32$ & $5.56$ & $132.31$ \\
\hline
\end{tabular}
\tablefoot{
List of simulations with differentiating parameters and some initial properties (at $\tphys=0$). Label identifies particular simulation and will be used throughout the paper. TF/nTF in the label means initially tidally filling or non-tidally filling. NF, when present, means that new features were included in this model (see text for details); \npop\ - number of stellar generations; \concpop\ - ratio between the \rh\ of the second and first stellar generation; \fracb\ - binary fraction; \rgc\ - galactocentric radius [\kpc]; \rh\ - half-mass radius for the first population [\pc]; \smt\ - total stellar mass [\msun]; \rc\ - core radius [\pc]; \xrc\ - core mass [\msun]; \roc\ - central density [$\msun / \pc^3$]; \threl\ - Spitzer half-mass relaxation time [\msun]. All simulations include the escapers information (see text). For all these simulations the counterpart field population (with dynamics turned off) was calculated (\emph{nodyn} in the label; not shown in the table) with the same parameters.
}
\end{table*}

\begin{table*}
\caption{\label{tab:ulx_group_fracions}Fraction of ULXs in different groups across simulations}
\centering
\begin{tabular}{lrrrrrr}
\hline\hline
 label & \gbhms & \gbhhg & \gbhcheb & \gimbh & \gns & other \\
\hline
npop1-fb10-TF & $0.17$ & $0.33$ & $0.36$ &  & $0.11$ & $0.04$ \\
npop1-fb10-nTF & $0.16$ & $0.32$ & $0.30$ &  & $0.12$ & $0.11$ \\
npop1-fb10-nTF-nodyn & $0.18$ & $0.36$ & $0.37$ &  & $0.05$ & $0.04$ \\
\\
npop1-fb95-TF & $0.11$ & $0.40$ & $0.33$ &  & $0.12$ & $0.04$ \\
npop1-fb95-nTF & $0.12$ & $0.39$ & $0.28$ &  & $0.11$ & $0.11$ \\
npop1-fb95-nTF-nodyn & $0.11$ & $0.43$ & $0.31$ &  & $0.11$ & $0.04$ \\
\\
npop2-cpop05-fb10-TF & $0.09$ & $0.39$ & $0.29$ &  & $0.09$ & $0.13$ \\
npop2-cpop05-fb10-nTF & $0.07$ & $0.29$ & $0.20$ & $0.33$ & $0.04$ & $0.06$ \\
npop2-cpop05-fb10-nTF-nodyn & $0.10$ & $0.46$ & $0.32$ &  & $0.04$ & $0.07$ \\
\\
npop2-cpop05-fb10-TF-NF & $0.15$ & $0.38$ & $0.27$ &  & $0.05$ & $0.14$ \\
npop2-cpop05-fb10-nTF-NF & $0.06$ & $0.25$ & $0.16$ & $0.44$ & $0.05$ & $0.03$ \\
npop2-cpop05-fb10-nTF-nodyn-NF & $0.11$ & $0.48$ & $0.33$ &  &  & $0.08$ \\
\\
npop2-cpop05-fb95-TF & $0.11$ & $0.29$ & $0.42$ &  & $0.11$ & $0.06$ \\
npop2-cpop05-fb95-nTF & $0.11$ & $0.25$ & $0.35$ & $0.16$ & $0.07$ & $0.06$ \\
npop2-cpop05-fb95-nTF-nodyn & $0.12$ & $0.29$ & $0.41$ &  & $0.14$ & $0.04$ \\
\\
npop2-cpop05-fb95-TF-NF & $0.13$ & $0.27$ & $0.44$ &  & $0.10$ & $0.06$ \\
npop2-cpop05-fb95-nTF-NF & $0.10$ & $0.18$ & $0.29$ & $0.34$ & $0.05$ & $0.04$ \\
npop2-cpop05-fb95-nTF-nodyn-NF & $0.12$ & $0.30$ & $0.43$ &  & $0.12$ & $0.04$ \\
\\
npop2-cpop2-fb10-TF & $0.10$ & $0.41$ & $0.30$ &  & $0.09$ & $0.10$ \\
npop2-cpop2-fb10-nTF & $0.06$ & $0.28$ & $0.19$ & $0.36$ & $0.04$ & $0.07$ \\
npop2-cpop2-fb10-nTF-nodyn & $0.10$ & $0.46$ & $0.32$ &  & $0.04$ & $0.07$ \\
\\
npop2-cpop2-fb10-TF-NF & $0.12$ & $0.39$ & $0.30$ &  & $0.09$ & $0.09$ \\
npop2-cpop2-fb10-nTF-NF & $0.16$ & $0.16$ & $0.09$ &  & $0.41$ & $0.18$ \\
npop2-cpop2-fb10-nTF-nodyn-NF & $0.11$ & $0.48$ & $0.33$ &  &  & $0.08$ \\
\\
npop2-cpop2-fb95-TF & $0.13$ & $0.26$ & $0.44$ &  & $0.13$ & $0.05$ \\
npop2-cpop2-fb95-nTF & $0.10$ & $0.24$ & $0.31$ & $0.20$ & $0.09$ & $0.06$ \\
npop2-cpop2-fb95-nTF-nodyn & $0.12$ & $0.29$ & $0.42$ &  & $0.13$ & $0.04$ \\
\\
npop2-cpop2-fb95-TF-NF & $0.12$ & $0.28$ & $0.45$ &  & $0.11$ & $0.05$ \\
npop2-cpop2-fb95-nTF-NF & $0.13$ & $0.17$ & $0.22$ &  & $0.34$ & $0.14$ \\
npop2-cpop2-fb95-nTF-nodyn-NF & $0.12$ & $0.29$ & $0.41$ &  & $0.14$ & $0.04$ \\
\hline
\end{tabular}
\tablefoot{
Fractional distribution of ULXs across different groups for various simulations. For group definitions see Table \ref{tab:ulx_groups}
}
\end{table*}
\end{appendix}

\label{lastpage}
\end{document}